\documentclass[a4paper,11pt]{article}
\usepackage{jinstpub} 
\usepackage{placeins}
\usepackage{float}
\usepackage{booktabs}
\usepackage{array}

\title{\boldmath Characterization of the RF Board for microwave SQUID multiplexing readout electronics}

\author[a]{J. Wang,}
\author[a]{J. Dong,}
\author[a]{Z. Cheng,}
\author[a,1]{X. Ren\note{Corresponding author.}}
\affiliation[a]{Key Laboratory of Particle Physics and Particle Irradiation (MOE),\\
Institute of Frontier and Interdisciplinary Science, Shandong University,\\
Qingdao, Shandong, 266237, China}

\emailAdd{renxx@sdu.edu.cn}

\abstract{Microwave SQUID multiplexing \((\mu\mathrm{MUX})\) is a widely used readout technique for large-scale transition-edge sensor (TES) arrays.
It uses radio-frequency (RF) probe tones to interrogate cryogenic resonators, requiring frequency conversion between the baseband electronics and the cryogenic RF signal chain.
This work describes the RF Board, a room-temperature frequency-conversion board deployed in the AliCPT \(\mu\mathrm{MUX}\) readout system.
The board up-converts \(0\)--\(4\)~GHz baseband I/Q signals to the \(4\)--\(8\)~GHz RF band for injection into the cryogenic chain and down-converts returned RF signals to baseband I/Q for ADC digitization.
For the 1000-tone design condition, with a DAC output tone power of \(-30\)~dBm/tone, the required power windows are \(-35\) to \(-25\)~dBm/tone for the RF tones transmitted to the cryostat and \(-45\) to \(-35\)~dBm/tone at the ADC input for the returned tones.
The RF Board is characterized using swept single-tone measurements covering up-conversion, down-conversion, and RF loopback, together with 560-tone measurements. Based on the single-tone measurements, the RF output power is calculated to be \(-31.06\) to \(-25.53\)~dBm, satisfying the RF output window.
Assuming a representative cryogenic-chain transmission of \(-40\)~dB, the loopback result gives an estimated returned power of \(-45\) to \(-35\)~dBm, within the target range.
A 560-tone measurement under the current AliCPT operating condition showed a stable linear response of the up-conversion path, with the identified in-band spurious components reduced below \(-80\)~dBm after spurious suppression and the spurious-free dynamic range (SFDR) exceeding \(43.10\)~dBc.
These results show that the RF Board meets the wideband frequency-conversion, tone-power, and multi-tone operation requirements of the \(\mu\mathrm{MUX}\) readout system.}


\begin{document}
\maketitle
\flushbottom

\section{Introduction}

Ground-based measurements of cosmic microwave background (CMB) polarization increasingly rely on large-scale arrays of transition-edge sensor (TES) bolometers \cite{Abazajian2016,Abitbol2017,Irwin2005,Ade2015}. As these arrays scale up, readout systems must support higher multiplexing density to reduce cryogenic wiring and thermal loading while preserving the required science performance \cite{Abitbol2017,Yu2023}. Microwave SQUID multiplexing \((\mu\mathrm{MUX})\) addresses this requirement by coupling TES bolometers, via RF-SQUIDs, to microwave resonators with unique resonance frequencies. Detector signals correspond to shifts in the resonator frequencies, which are read out using probe tones transmitted through a common RF feedline \cite{Mates2008,Dober2021,Becker2019}. This approach has been employed for on-sky observations with the Keck Array and adopted for large-scale deployment in the Simons Observatory \cite{Cukierman2020,McCarrick2021,Satterthwaite2024}.

\begin{figure}[!htbp]
\centering
\includegraphics[width=0.86\textwidth]{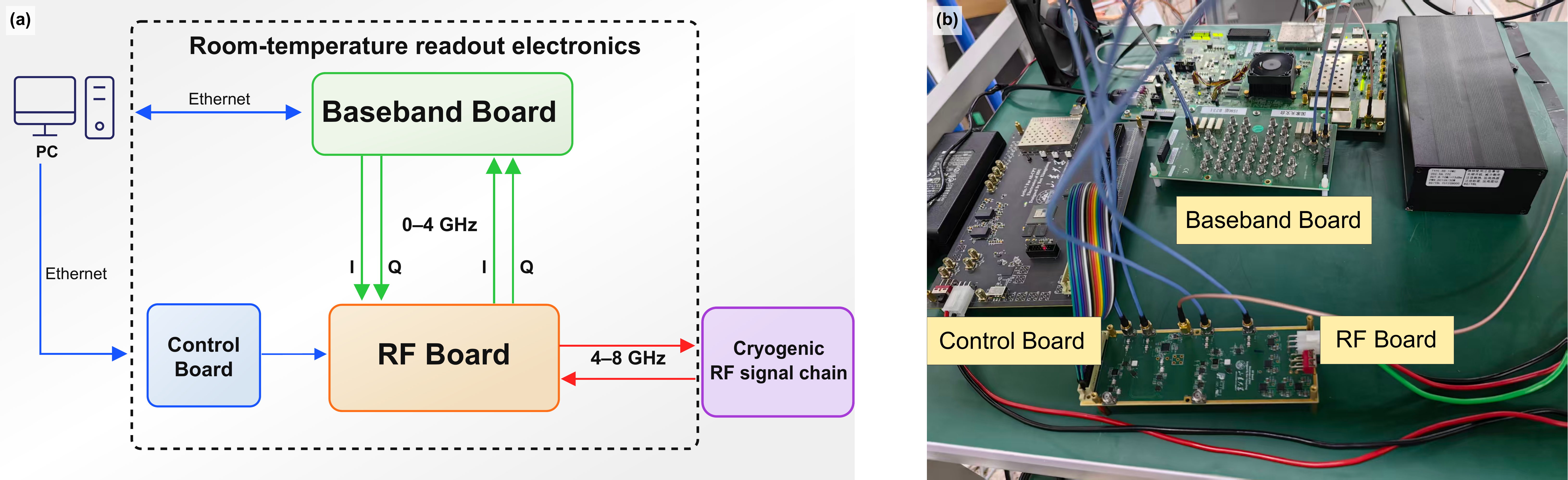}
\caption{Room-temperature electronics in the \(\mu\mathrm{MUX}\) readout chain. (a) System block diagram of the Baseband Board, Control Board, and RF Board. (b) Photograph of the electronics system.}
\label{fig:room_temperature_electronics}
\end{figure}

The cryogenic resonators are read out by room-temperature electronics that generate probe tones and acquire the returned signals \cite{Gard2018}. The room-temperature electronics deployed in the AliCPT \(\mu\mathrm{MUX}\) readout chain consist of a Baseband Board, a Control Board, and an RF Board, as shown in Figure~\ref{fig:room_temperature_electronics} \cite{Henderson2018}. The Baseband Board, based on a Xilinx ZCU111 RFSoC evaluation board, generates and acquires baseband I/Q signals. The Control Board configures the RF Board. The RF Board serves as the bridge linking the Baseband Board and the cryogenic RF signal chain. It provides bidirectional frequency conversion between the \(0\)--\(4\)~GHz baseband I/Q signals and the \(4\)--\(8\)~GHz RF band.

In this role, the RF Board sets the RF output power transmitted to the cryostat and the returned power sent to the ADC. These power levels are key parameters in the readout power budget, as they influence detector noise performance and readout signal-to-noise ratio (SNR). As the electronics noise floor is approximately independent of the readout-tone power, the tone power should be maximized within the linear regime of the system to improve the readout SNR. For the nominal 1000-tone design condition evaluated here, a frequency comb spans the \(4\)~GHz bandwidth with a baseline DAC output of \(-30\)~dBm/tone. After up-conversion, the RF tone power must be maintained within a \(-35\) to \(-25\)~dBm/tone window. This window is set to ensure that, after approximately \(45\)~dB of cold attenuation, the probe-tone power at the resonator inputs falls within the optimal range of \(-80\) to \(-70\)~dBm/tone \cite{Dober2021,Dober2017,Dutcher2024}. On the return path, the tones pass through cold amplifiers with a gain of about \(25\)~dB before entering the RF Board's receive chain, which is designed to convert and transmit the returned tones to the ADC input within a target range of \(-45\) to \(-35\)~dBm/tone. The main operating targets and characterization settings are summarized in Table~\ref{tab:operating_targets}.

\begin{table}[!htbp]
\centering
\caption{Main operating targets and characterization settings for the RF Board.}
\label{tab:operating_targets}
\smallskip
\footnotesize
\setlength{\tabcolsep}{5pt}
\renewcommand{\arraystretch}{1.288}
\begin{tabular}{@{}>{\raggedright\arraybackslash}p{0.30\textwidth}>{\raggedright\arraybackslash}p{0.23\textwidth}>{\raggedright\arraybackslash}p{0.42\textwidth}@{}}
\toprule
Parameter & Value & Basis in this work\\
\midrule
\(\mu\mathrm{MUX}\) RF band & \(4\)--\(8\)~GHz & Operating RF band of the cryogenic \(\mu\mathrm{MUX}\) resonators\\
Baseband I/Q band & \(0\)--\(4\)~GHz & Baseband frequency range supported by the DAC and ADC on the Baseband Board\\
Designed multiplexing factor & 1000:1 & Design target of the \(\mu\mathrm{MUX}\) readout architecture\\
Operating multiplexing factor & 560:1 & Multiplexing factor used in the current \(\mu\mathrm{MUX}\) readout condition\\
DAC full-scale output power & \(+5\)~dBm & Typical full-scale output power of the ZCU111 RF-DAC\\
DAC output tone power & \(-30\)~dBm/tone & Representative per-tone output power selected for the 1000-tone design target to use the available DAC dynamic range\\
DAC sampling rate & 4.096~GSa/s & Sampling rate configured for the ZCU111 RF-DAC in this work\\
ADC full-scale input power & \(+1\)~dBm & Typical full-scale input power of the ZCU111 RF-ADC\\
ADC sampling rate & 4.096~GSa/s & Sampling rate configured for the ZCU111 RF-ADC in this work\\
Optimal resonator input tone power & \(-80\) to \(-70\)~dBm/tone & Optimal probe-tone power at the \(\mu\mathrm{MUX}\) resonator inputs\\
Required up-converted RF tone power & \(-35\) to \(-25\)~dBm/tone & Required RF tone power transmitted to the cryostat\\
Target ADC input tone power & \(-45\) to \(-35\)~dBm/tone & Target per-tone input power range selected for the 1000-tone design target to use the available ADC dynamic range\\
Characterization frequency range & \(10\)~MHz--\(4\)~GHz baseband; \(4.01\)--\(8\)~GHz RF & Set by the VNA minimum frequency of \(10\)~MHz\\
Characterization frequency step & \(7.8125\)~kHz & Minimum frequency spacing supported by the Baseband Board\\
\bottomrule
\end{tabular}
\end{table}

\FloatBarrier

\pagebreak[4]

To verify these performance targets, swept single-tone measurements were performed on the RF Board to characterize the up-conversion, down-conversion, and RF loopback responses. Specifically, the up- and down-conversion measurements were used to quantify the RF output power and the receive-path gain response, respectively. The RF loopback measurement was then used to establish a board-level bidirectional reference for estimating the returned signal power at the ADC input. In addition, 560-tone measurements corresponding to the current AliCPT operating condition were performed to evaluate the spectral performance and linearity of the up-conversion chain under practical \(\mu\mathrm{MUX}\) operation.

\FloatBarrier

\section{RF Board Architecture}

\begin{figure}[!htbp]
\centering
\includegraphics[width=0.92\textwidth]{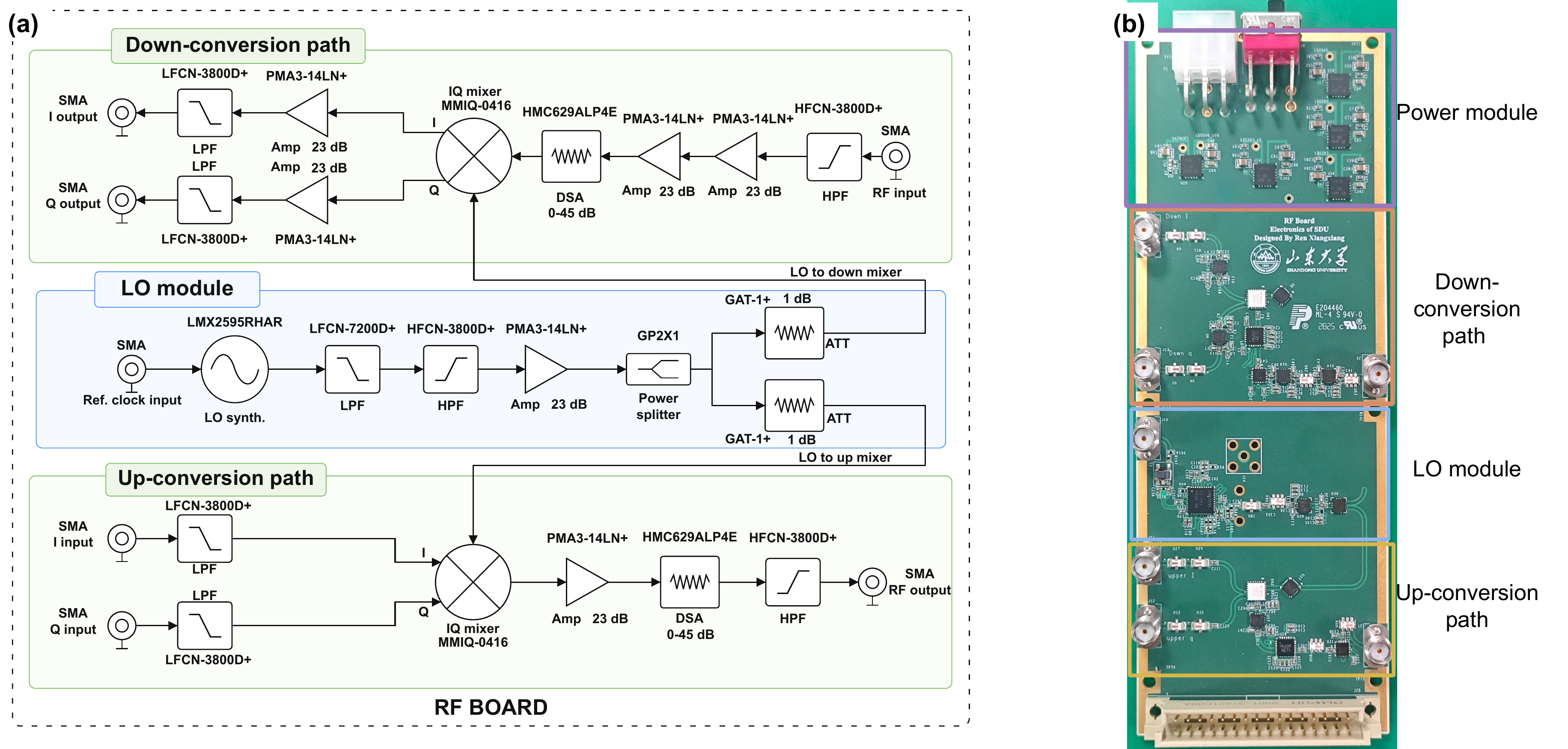}
\caption{RF Board architecture and hardware implementation. (a) Functional block diagram showing the up-conversion path, down-conversion path, and LO module. (b) Photograph of the RF Board.}
\label{fig:rf_board_architecture}
\end{figure}

Functioning as the bidirectional translation interface between the baseband electronics and the cryogenic RF chain, the RF Board translates the \(0\)--\(4\)~GHz baseband I/Q signals to the \(4\)--\(8\)~GHz RF band to drive the cryogenic resonators, while converting the returned tones back to baseband for ADC acquisition. Figure~\ref{fig:rf_board_architecture} shows the architecture and hardware implementation of the RF Board. The board is organized into four functional blocks: the local oscillator (LO) module, the up-conversion path, the down-conversion path, and the power module. The LO module generates and distributes a common LO signal to both conversion paths and accepts an external reference input, while the power module supplies the voltages for the on-board components.

In the LO module, the shared LO signal is generated by a Texas Instruments LMX2595 frequency synthesizer. An external reference clock is provided to the LMX2595 through the Ref. clock input interface for frequency synchronization. The LO signal is filtered by Mini-Circuits LFCN-7200D+ and HFCN-3800D+ filters to suppress unwanted harmonic components and spurious signals, and then amplified by a PMA3-14LN+ amplifier (23~dB gain) before being distributed to the two Marki
Microwave MMIQ-0416LSM IQ mixers through a GP2X1 power splitter.

In the up-conversion path, the two I/Q signals generated by the DACs are first filtered by LFCN-3800D+ low-pass filters to suppress high-frequency spurious components from the DACs. The filtered I/Q signals are then up-converted to the RF band by the mixer using the shared LO signal. The converted RF signal is subsequently amplified by a PMA3-14LN+ amplifier, and its power is adjusted by an HMC629ALP4E digital step attenuator. Finally, HFCN-3800D+ high-pass filters are applied to suppress the unwanted lower sideband generated by the mixer, and the resulting RF signal is delivered to the cryogenic RF chain.

In the down-conversion path, the returned RF signal is first filtered by HFCN-3800D+ high-pass filters. Two cascaded PMA3-14LN+ amplifiers are employed as the returned RF signal power is relatively low. The signal is then passed through the digital step attenuator and down-converted into I/Q signals by the IQ mixer. After additional amplification and low-pass filtering to remove high-frequency mixing products, the recovered I/Q signals are delivered to the ADCs. In both the LO module and the signal paths, the soldered filters can be replaced with different models to accommodate different LO frequencies and signal bandwidths.

\section{RF Board Performance}

To evaluate the performance of the RF Board, we performed both swept single-tone measurements and multi-tone measurements under representative AliCPT operating conditions. The swept single-tone measurements were carried out in three configurations: up-conversion, down-conversion, and RF loopback. For the swept single-tone measurements, the conversion and loopback responses are reported as effective \(S_{21}\) magnitudes in dB, calculated from the corresponding output and input powers. The up-conversion measurements tested the transmit path and were used to calculate the RF output power. The down-conversion measurements characterized the receive-path response. The RF loopback measurements provided the combined bidirectional response of the RF Board, which was used as a board-level reference for estimating the returned power. The multi-tone measurements evaluate the spectral performance and up-conversion linearity of the RF Board under practical operation.

\begin{figure}[!htbp]
\centering
\includegraphics[width=0.64\textwidth]{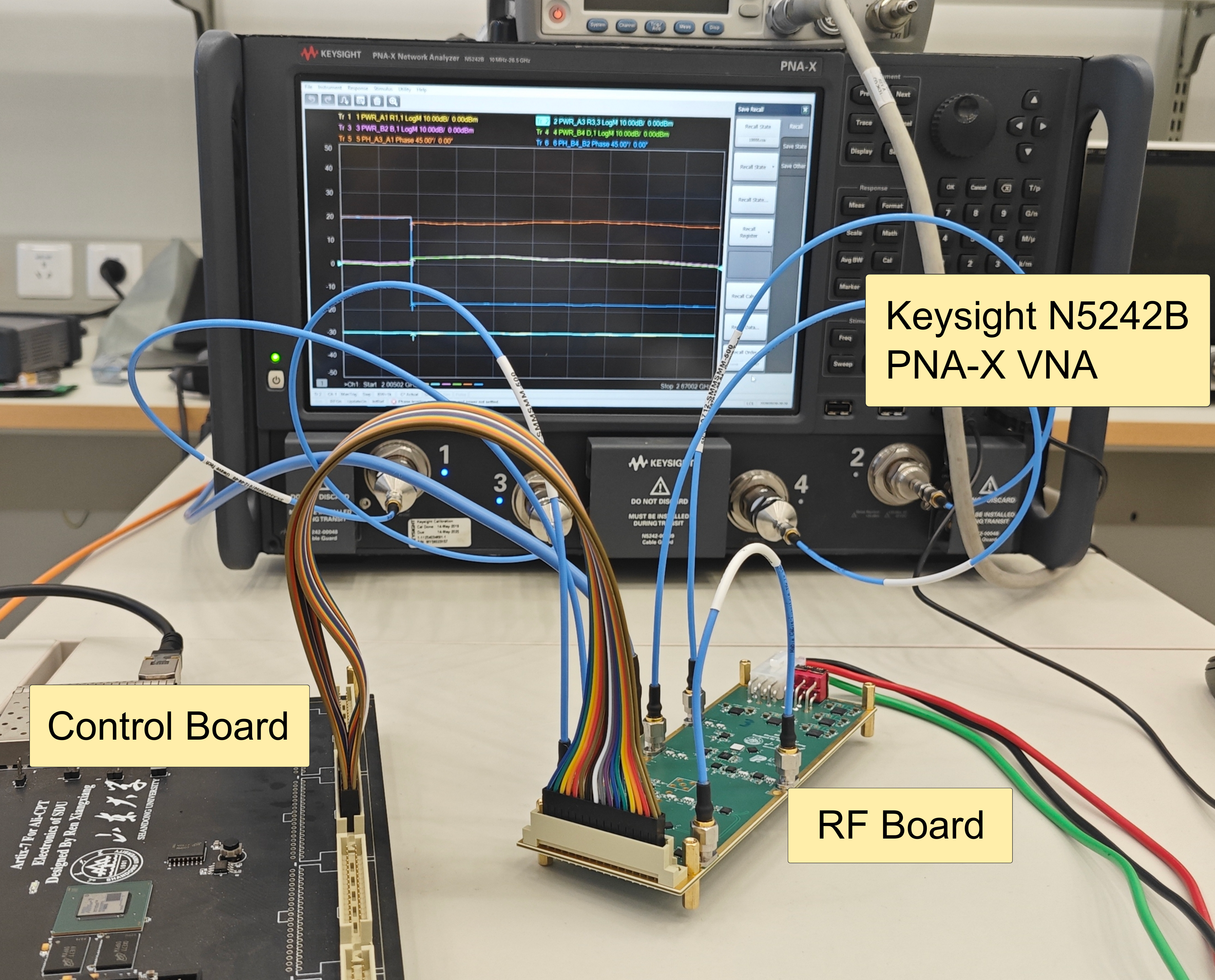}
\caption{Measurement setup for the swept single-tone characterization, including the Keysight N5242B PNA-X vector network analyzer, the RF Board under test, and the Control Board used to configure the RF Board.}
\label{fig:measurement_setup}
\end{figure}

All swept single-tone measurements were carried out using a Keysight N5242B PNA-X vector network analyzer. The RF Board LO was set to \(4\)~GHz and locked to the \(10\)~MHz reference output of the VNA. Before data acquisition, the VNA was calibrated using a Keysight N4433D electronic calibration module and a Keysight U2000A power meter. This calibration corrected the test-cable losses and the relative delay between the I and Q paths, thereby moving the measurement reference planes to the cable ends connected to the RF Board. It also ensured that the swept baseband I/Q tones applied to the RF Board inputs maintained the required \(90^\circ\) phase relationship, consistent with the quadrature I/Q drive used in operation. The frequency sweeps covered \(10\)~MHz--\(4\)~GHz at baseband and \(4.01\)--\(8\)~GHz in the RF band, with a frequency step of \(7.8125\)~kHz, matching the minimum frequency spacing supported by the Baseband Board. The characterization setup for the swept single-tone measurements is shown in Figure~\ref{fig:measurement_setup}. For the multi-tone measurement, the DACs on the Baseband Board generated the frequency comb, which was then applied to the RF Board input. The resulting RF output spectrum was measured using a R\&S FPL1026 spectrum analyzer.

\subsection{Up-conversion response and RF output power}

The up-conversion measurements were used to verify whether the transmit path provides the required RF power transmitted to the cryostat. In the 1000-tone design configuration of the \(\mu\mathrm{MUX}\) readout system, the DACs on the Baseband Board generate 1000 baseband I/Q tones, each with an output power of \(-30\)~dBm/tone. The RF Board then up-converts these tones to the \(4\)--\(8\)~GHz RF band and transmits them to the cryogenic chain. To reproduce this condition, the VNA supplied two swept single-tone signals with a \(90^\circ\) relative phase offset to the I and Q inputs of the RF Board, with each input power set to \(-30\)~dBm/tone. For this path, the effective \(S_{21}\) magnitude was derived from the RF output power and the combined baseband I/Q input power, with the programmable attenuator set to three values used during laboratory operation: \(6\), \(9\), and \(15\)~dB. At a DAC output power of \(-30\)~dBm/tone, the nominal programmable attenuation setting is \(9\)~dB, while the \(6\)~dB and \(15\)~dB settings correspond to other DAC output tone-power levels.

\begin{figure}[!htbp]
\centering
\includegraphics[width=0.94\textwidth]{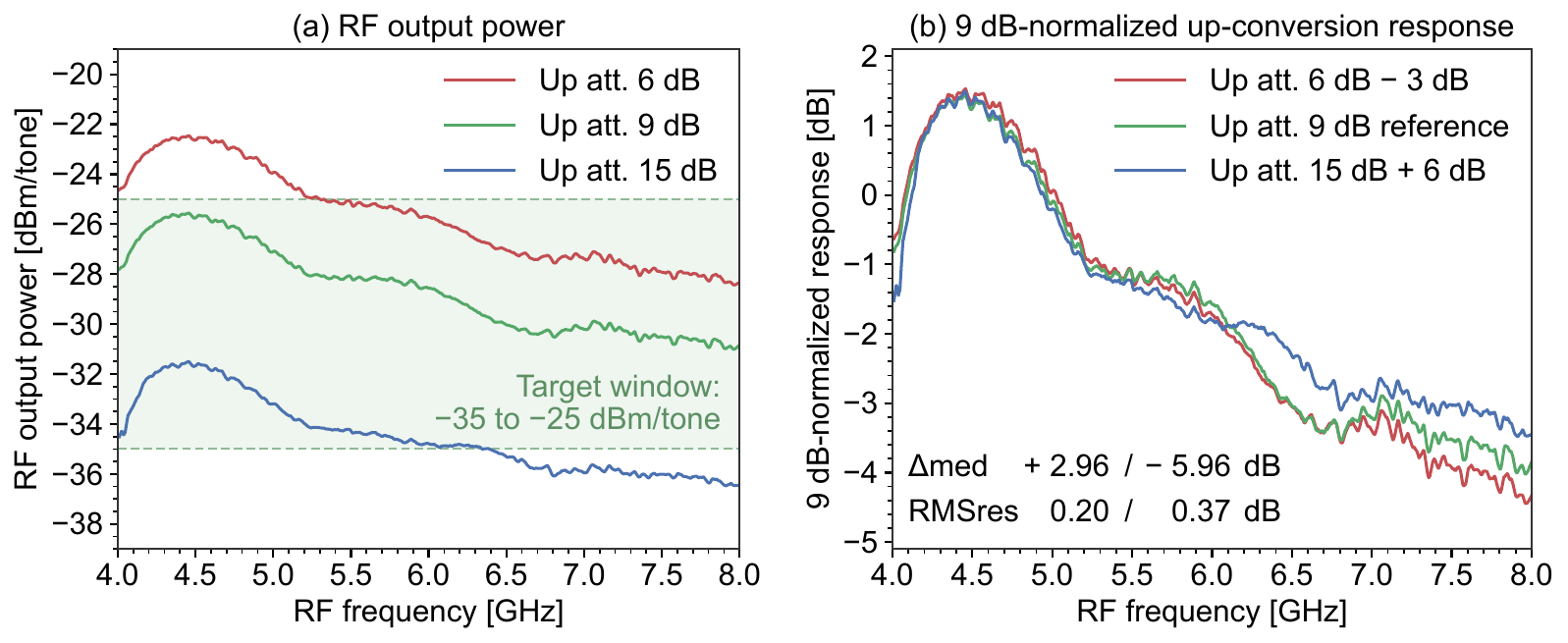}
\caption{Up-conversion characterization. (a) RF output power calculated at the three attenuation settings. The shaded region indicates the required RF output power window of \(-35\) to \(-25\)~dBm/tone. (b) Offset-normalized up-conversion traces referenced to the \(9\)~dB attenuation setting.}
\label{fig:upconversion}
\end{figure}

The up-conversion response was then used to calculate the RF output power. Using the combined baseband I/Q input power of \(-27\)~dBm and the measured \(S_{21}\) magnitude range of \(-4.06\) to \(1.47\)~dB at the \(9\)~dB setting, the RF output power is calculated to be \(-31.06\) to \(-25.53\)~dBm over the operating band, as shown in Figure~\ref{fig:upconversion}(a). This range falls within the required \(-35\) to \(-25\)~dBm/tone window summarized in Table~\ref{tab:operating_targets}. Relative to this setting, the \(6\)~dB setting increases the RF output power, whereas the \(15\)~dB setting lowers it, providing a tuning range for determining the optimal drive power for the cryogenic resonators.

The traces at different attenuation settings further indicate that the programmable attenuator primarily shifts the overall level while leaving the wideband shape largely unchanged. To verify the interpretation, the \(6\)~dB and \(15\)~dB curves were shifted by \(-3\)~dB and \(+6\)~dB, respectively, to align them with the \(9\)~dB reference trace. After this normalization, the three curves closely overlap across most of the frequency band, as shown in Figure~\ref{fig:upconversion}(b). The median shifts from the \(9\)~dB reference are \(+2.96\)~dB and \(-5.96\)~dB for the \(6\)~dB and \(15\)~dB traces, respectively, with corresponding RMS residuals of \(0.20\)~dB and \(0.37\)~dB. These values indicate that the programmable attenuator provides predictable gain adjustment while largely preserving the frequency-dependent response. During system operation, the measured curve could also be used to pre-equalize the DAC tone powers, thereby reducing frequency-dependent variations in the RF output power.

\FloatBarrier

\subsection{Down-conversion response}

The receive-path conversion response was evaluated using down-conversion measurements. The VNA supplied a swept RF tone to the board's RF input, and the resulting baseband I/Q outputs were recorded. The down-conversion \(S_{21}\) was then derived by comparing the combined baseband I/Q output power against the applied RF input power. These tests were performed at three representative receive-path attenuation settings: \(21\), \(18\), and \(15\)~dB.

\begin{figure}[!htbp]
\centering
\includegraphics[width=0.94\textwidth]{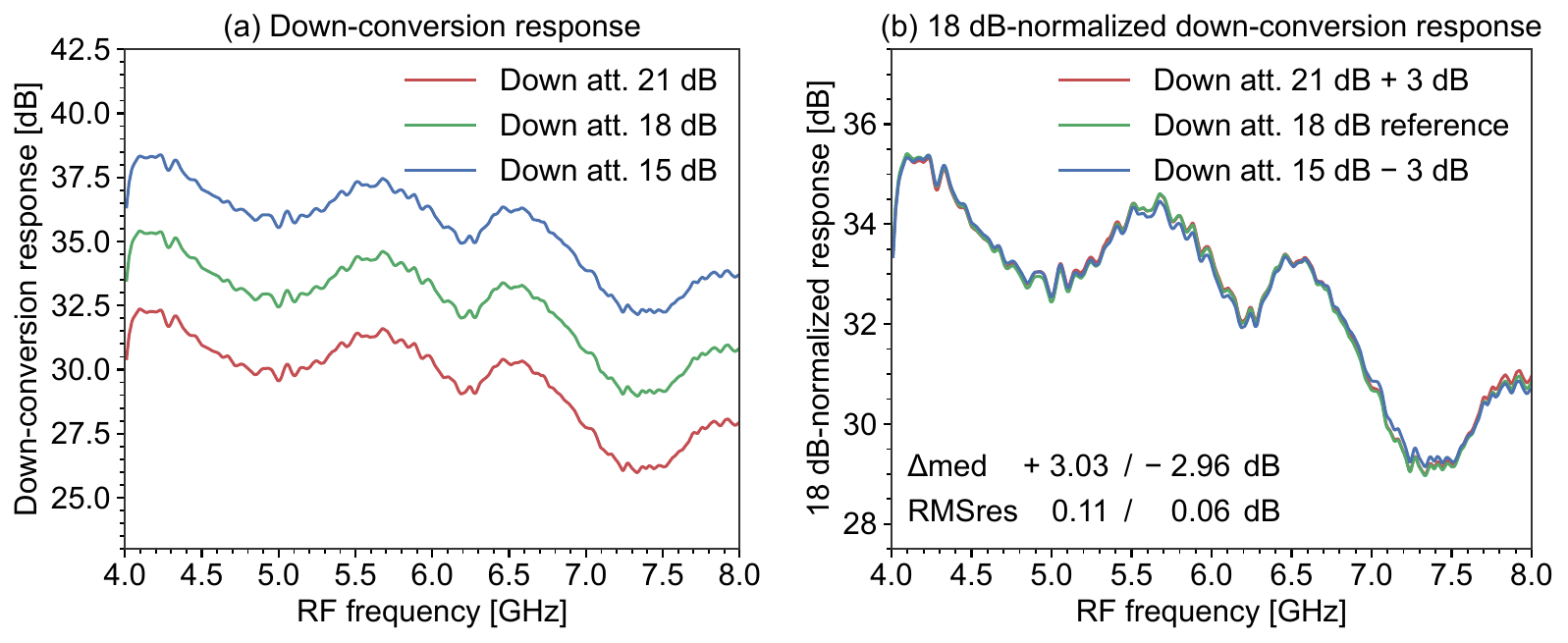}
\caption{Down-conversion characterization. (a) Down-conversion traces at three programmable attenuation settings. (b) Offset-normalized down-conversion traces referenced to the \(18\)~dB attenuation setting.}
\label{fig:downconversion}
\end{figure}

Figure~\ref{fig:downconversion}(a) shows the expected dependence on the attenuation setting. Following the same normalization procedure used in Section~3.1, the three curves closely overlap, as shown in Figure~\ref{fig:downconversion}(b). The \(15\)~dB trace relative to the \(18\)~dB reference gives a median shift of \(+3.03\)~dB with an RMS residual of \(0.11\)~dB, while the \(21\)~dB trace gives a median shift of \(-2.96\)~dB with an RMS residual of \(0.06\)~dB. The measured shifts are consistent with the expected attenuation changes set by the receive-path programmable attenuator.

\FloatBarrier

\subsection{RF loopback response}

\begin{figure}[H]
\centering
\includegraphics[width=0.84\textwidth]{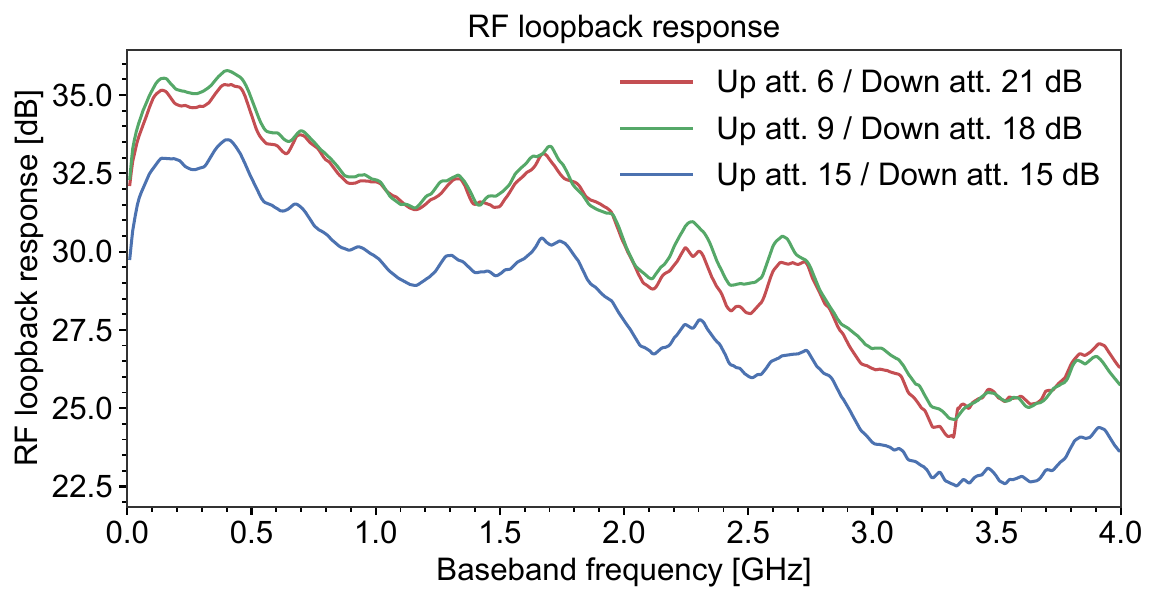}
\caption{RF loopback responses at up/down attenuation settings of \(6/21\), \(9/18\), and \(15/15\)~dB.}
\label{fig:rf_loopback}
\end{figure}

The RF loopback measurements were used to characterize the combined behavior of the transmit and receive paths of the RF Board. For this test, the up-conversion RF output was connected directly to the down-conversion RF input. Figure~\ref{fig:rf_loopback} compares the RF loopback results for three pairs of up- and down-conversion attenuation settings: \(6/21\), \(9/18\), and \(15/15\)~dB. The two \(27\)~dB total-attenuation cases nearly overlap, whereas the \(30\)~dB case shows a lower overall level.

\begin{figure}[H]
\centering
\includegraphics[width=0.92\textwidth]{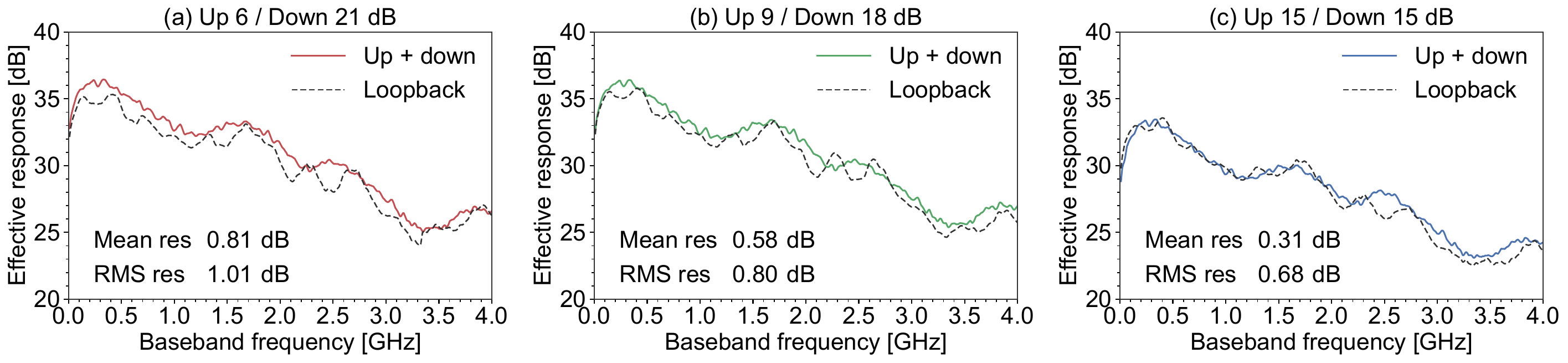}
\caption{Comparison of the summed independently obtained up- and down-conversion curves with the directly measured RF loopback responses at up/down attenuation settings of (a) \(6/21\)~dB, (b) \(9/18\)~dB, and (c) \(15/15\)~dB. The insets give the mean and RMS residuals.}
\label{fig:loopback_comparison}
\end{figure}

The direct loopback result was then compared with a reconstructed result calculated as the sum of the independently obtained up- and down-conversion responses. As shown in Figure~\ref{fig:loopback_comparison}, the two methods match closely for all three setting pairs. The mean residuals are \(0.81\), \(0.58\), and \(0.31\)~dB, with RMS residuals of \(1.01\), \(0.80\), and \(0.68\)~dB for the \(6/21\), \(9/18\), and \(15/15\)~dB setting pairs, respectively. The residual statistics indicate negligible discrepancy between the two approaches, validating the direct loopback result as the board-level reference.

After accounting for the cryogenic-chain transmission, the board-level test results can be used to evaluate the power budget at the ADC input. The cryogenic detector chain typically has a transmission of \(-20\) to \(-40\)~dB, and \(-40\)~dB was adopted as a representative value for this estimate. For the representative \(9/18\)~dB setting pair shown in Figure~\ref{fig:loopback_comparison}(b), the board-level loopback \(S_{21}\) magnitude ranges from \(25\) to \(35\)~dB. When combined with the \(40\)~dB transmission loss of the cryogenic chain, the total system \(S_{21}\) translates to \(-15\) to \(-5\)~dB. With a DAC output tone power of \(-30\)~dBm/tone, the corresponding power at the ADC input is \(-45\) to \(-35\)~dBm/tone per channel, within the target range shown in Table~\ref{tab:operating_targets}. For cryogenic chains with lower attenuation, the RF Board's programmable attenuation can be increased accordingly. As discussed in Section~3.1, pre-equalizing the DAC tone powers could further reduce variations in the returned power.

Furthermore, the validated loopback \(S_{21}\) can be used as a reference to derive the effective \(S_{21}\) of the cryogenic chain. Once the cryogenic chain is connected, the measured full-link response will include both the RF Board contribution and the cryogenic-chain transmission. The effective cryogenic-chain \(S_{21}\) can then be obtained by subtracting the loopback \(S_{21}\) from the full-link response.

\subsection{Multi-tone response}

\begin{figure}[H]
\centering
\includegraphics[width=0.94\textwidth]{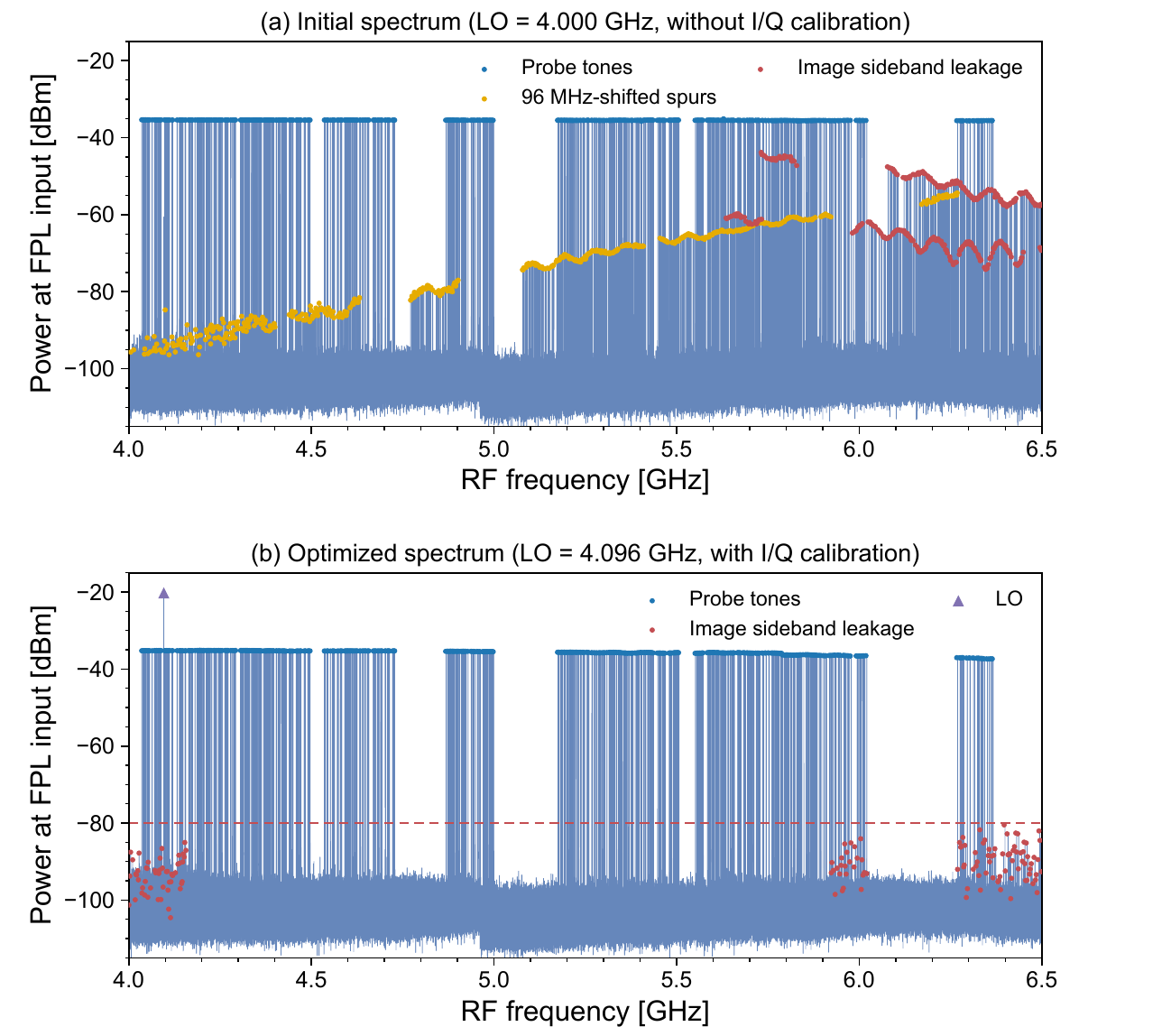}
\caption{RF output spectra under the 560-tone operating condition: (a) before spurious suppression; (b) after LO-frequency optimization and I/Q calibration.}
\label{fig:multitone_spectrum}
\end{figure}

To evaluate the RF Board under realistic multi-tone operating conditions, 560-tone measurements were performed using the tone configuration currently adopted in the AliCPT readout system, with tones distributed across the \(4.035\)--\(6.364\)~GHz RF band. The 560-tone frequency combs were generated by the DACs on the Baseband Board, up-converted by the RF Board, and measured at the RF output using an R\&S FPL1026 spectrum analyzer. The RF output spectra were acquired with a resolution bandwidth (RBW) of \(1\)~kHz, except for the region near the LO in Figure~\ref{fig:multitone_spectrum}(b), where an RBW of \(100\)~Hz was used to better resolve LO-related spurious sidelobes.

During the multi-tone measurement, a group of in-band spurious components was observed, appearing as a frequency-comb replica shifted by \(96\)~MHz relative to the target multi-tone signals. This frequency offset was found to be related to the difference between the LO frequency and the DAC sampling rate. With the LO frequency set to \(4\)~GHz and the DAC sampling rate at \(4.096\)~GSa/s, the \(96\)~MHz frequency mismatch caused the mixing products between the DAC sampling images and the second harmonic of the LO to fall at the observed offset position. To suppress this systematic in-band spur, the LO frequency was adjusted to \(4.096\)~GHz to match the DAC sampling rate, causing these mixing products to overlap with the desired tones rather than appearing as independent spurious components. After the LO frequency adjustment, RF tones located on both sides of the LO frequency were generated by assigning the corresponding positive and negative frequency components in the digital I/Q waveform, maintaining coverage of the intended RF operating range.

The remaining image sideband leakage was mainly caused by amplitude and phase imbalance in the I/Q conversion chain. Ideally, I/Q up-conversion requires the I and Q signals to have identical amplitudes and an exact \(90^\circ\) phase difference. However, practical imperfections in the I/Q conversion chain introduce amplitude and phase imbalance, reducing the cancellation of the unwanted sidebands. Therefore, I/Q calibration was performed by compensating the relative amplitude and phase of the I/Q waveforms. The compensation parameters were optimized by minimizing the image power in the RF output spectrum.

Figure~\ref{fig:multitone_spectrum} shows the RF output spectra before and after the spurious suppression process. The observed \(96\)~MHz-shifted frequency-comb replica and image-related spurious components were effectively suppressed after LO optimization and I/Q calibration, with all identified in-band spurious components reduced below \(-80\)~dBm. The SFDR of the RF Board under the 560-tone operating condition was conservatively calculated to be better than \(43.10\)~dBc, where SFDR is defined here as the power difference between the weakest tone and the highest spurious component.

\begin{figure}[!htbp]
\centering
\includegraphics[width=0.94\textwidth]{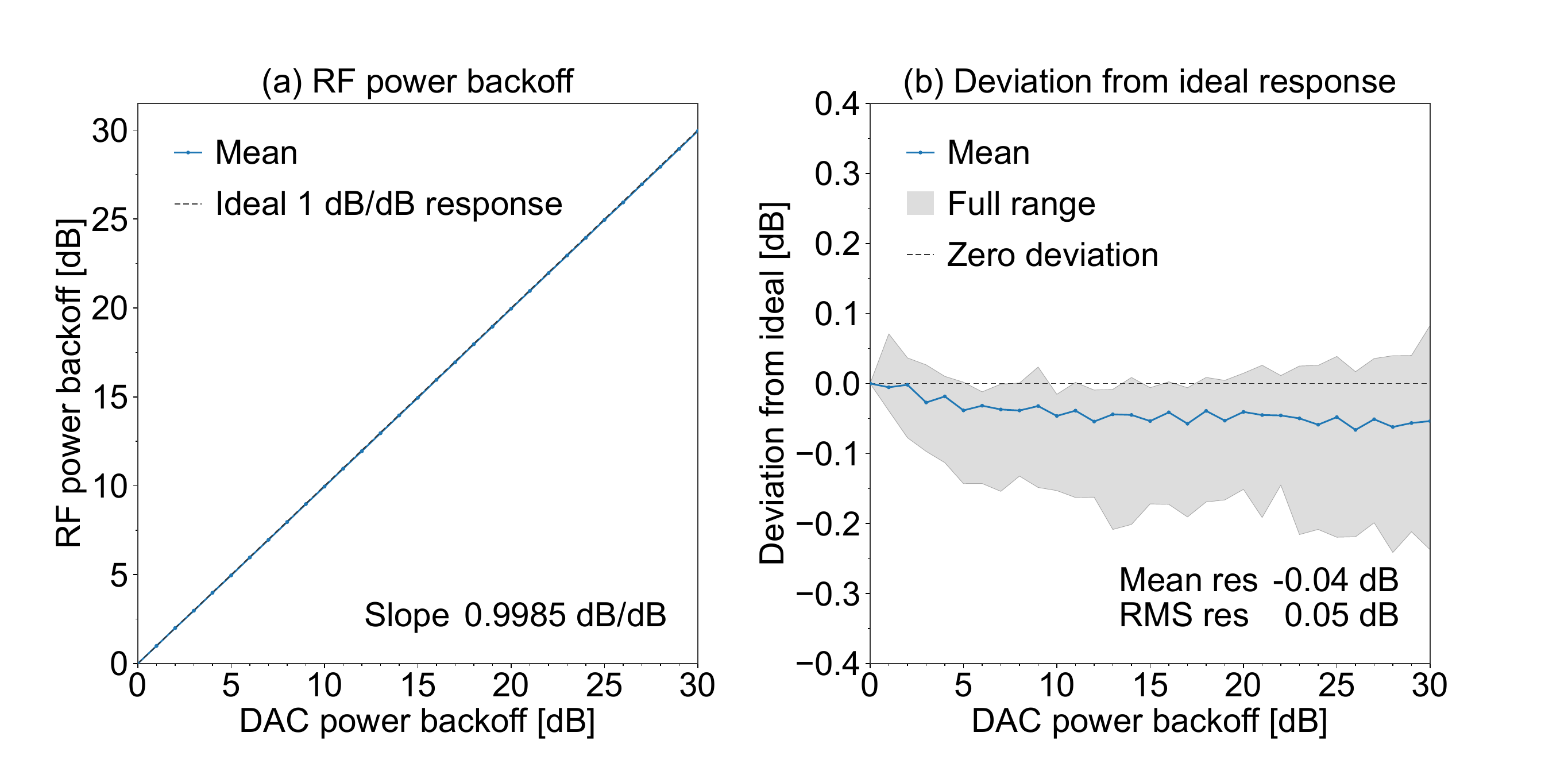}
\caption{Up-conversion path linearity under the 560-tone operating condition. (a) Mean RF power backoff compared with the ideal 1~dB/dB response. (b) Deviations of the measured RF power backoff from the ideal DAC backoff. The blue curve and gray band indicate the mean and full range across tones, respectively.}
\label{fig:multitone_linearity}
\end{figure}

The linearity of the up-conversion path was further evaluated under the 560-tone operating condition. Starting from the DAC power configuration used for this measurement, the excitation was reduced in 1~dB steps over a 30~dB range, with the I and Q waveform amplitudes scaled by the same factor. As shown in Figure~\ref{fig:multitone_linearity}(a), the mean RF power backoff of the 560 tones closely follows the applied DAC power backoff, with a fitted slope of 0.9985~dB/dB. In Figure~\ref{fig:multitone_linearity}(b), the deviation of each tone was calculated as its measured RF power backoff minus the ideal 1~dB/dB backoff. Across all tones and settings, the mean and RMS deviations were \(-0.04\) and 0.05~dB, respectively, and the maximum absolute deviation was 0.24~dB. These results indicate that the up-conversion path exhibits an approximately linear power response over the tested 30~dB range, with no appreciable gain compression observed under this configuration.

\FloatBarrier

\section{Conclusion}

This work has described the RF Board used as the room-temperature bidirectional frequency conversion interface in the \(\mu\mathrm{MUX}\) readout chain. Its performance was evaluated using swept single-tone measurements of up-conversion, down-conversion, and RF loopback, together with 560-tone measurements.

The up-conversion and down-conversion measurements established the effective \(S_{21}\) of the transmit and receive paths, respectively. For both paths, the attenuation settings produced the expected gain shifts while largely preserving the wideband shapes. The RF output power calculated for the 1000-tone design target ranged from \(-31.06\) to \(-25.53\)~dBm, meeting the required power specification of \(-35\) to \(-25\)~dBm/tone. The RF loopback measurement was shown to provide a valid board-level characterization and was used to estimate a returned power of approximately \(-45\) to \(-35\)~dBm, which falls within the target range.

In the 560-tone measurement, adjusting the LO frequency from \(4.000\) to \(4.096\)~GHz removed the distinct \(96\)~MHz-shifted frequency-comb replica, while I/Q calibration suppressed the remaining image-related spurious components. After these adjustments, all identified in-band spurious components were below \(-80\)~dBm, the SFDR was conservatively calculated to be better than \(43.10\)~dBc, and the up-conversion chain maintained a stable linear response with no appreciable gain compression over the tested power range. Together with the swept single-tone results, these findings indicate that the RF Board meets the requirements for wideband bidirectional frequency conversion and multi-tone operation, while remaining compatible with the power budget of the \(\mu\mathrm{MUX}\) readout system.

\acknowledgments

This work is supported by the National Key Research and Development Program of China (Grant No. 2022YFC2204904, No. 2020YFC2201704).

\bibliographystyle{JHEP}
\bibliography{biblio}
\end{document}